\documentclass[12pt]{article}
\usepackage{epsfig}
\usepackage{graphicx}

\def\beq{\begin{equation}}
\def\eeq#1{\label{#1}\end{equation}}
\def\eeqn{\end{equation}}

\def\beqa{\begin{eqnarray}}
\def\eeqa#1{\label{#1}\end{eqnarray}}
\def\eeqan{\end{eqnarray}}

\let\bar=\overbar

\def\Dslash{\not{\hbox{\kern-4pt $D$}}}
\def\dslash{\not{\hbox{\kern-2pt $\del$}}}

\def\msb{{\bar{\ssstyle M \kern -1pt S}}}

\usepackage{fancyhdr,graphicx}
\def\Title#1{\begin{center} {\Large {\bf #1} } \end{center}}

\begin{document}

\Title{The Nuclear Symmetry Energy in Heavy Ion Collisions}

\bigskip


\begin{raggedright}

{\it Hermann Wolter}\\
\bigskip
{\it Fakult\"at f\"ur Physik, Ludwig-Maximilians-Universit\"at M\"unchen,
85748 Garching}\\

\end{raggedright}

\bigskip

\centerline{\bf Abstract}
In this contribution I discuss the nuclear symmetry energy in the regime of hadronic degrees of freedom. 
The density dependence of the symmetry energy is important from very low
densities in supernova explosions, to the structure of neutron-rich nuclei
around saturation density, and to several times saturation density in neutron
stars. Heavy ion collisions are the only means to study this density dependence
in the laboratory. Numerical simulations of transport theories
are used to extract the equation-of-state, and thus also the symmetry energy. I discuss some
examples, which relate particularly to the high density symmetry energy, which is of particular interest today.  I review the status and point out some open problems in the  determination of the symmetry energy in heavy ion collisions.

\subsection*{1 Introduction}

The nuclear Equation-of-State (EoS) is often taken to specify the energy of nuclear matter as a function of density, temperature and asymmetry. For zero temperature it can be written as 
$E(\rho,\delta)=E_{nm}(\rho)+E_{sym}(\rho)\delta^2+...$, where $\rho$ is the total density of the system and $\delta=(\rho_n-\rho_p)/\rho$ the asymmetry with the neutron and proton densities. The term proportional to $\delta^2$ is the symmetry energy. For saturation density it is related to the symmetry energy term in the empirical mass formula. However, the dependence on density is of great importance in nuclei away from stability and in astrophysics in core-collapse supernovae and neutron stars, which have a large neutron excess and where a large range of densities from very low (in supernovae) to several times saturation density (in neutron stars) is involved. Predictions of mircroscopic many-body calculations of the symmetry energy differ widely, especially above saturation density \cite{Fuchs06}. Most likely the reason is the poorly known short range isovector repulsion \cite{BALi10}. Thus there are extensive efforts to determine the symmetry energy from nuclear structure, heavy ion collisions and astrophysical observations.
  
In nuclear structure one explores the symmetry energy around saturation density $\rho_0$, using an expansion of the form $E_{sym}=S_0+(L/3)(\rho-\rho_0)/\rho+...$, or a parametrization which splits the symmetry energy into a kinetic contribution taken in the form of the free Fermi gas, 
and a potential part as a power law $E_{sym}=\frac{1}{3}\epsilon_F (\rho/\rho_0)^{2/3}+C (\rho/\rho_0)^{\gamma}$ with a parameter $\gamma$. The slope of the symmetry energy $L$, or more specifically the correlation between the value and the slope, $S_0\, {\rm vs.}\, L$, has been extensively investigated using various observables, like nuclear masses, Giant and Pygmy dipole resonances, dipole polarizabilities, neutron skin radii (difference between neutron and proton radii), and isobaric analog state energies \cite{Tsang12,Horow14}. Neutron star observations, on the other hand, also provide access to the nuclear EoS. since it determines uniquely the mass-radius relation. The simultaneous measurement of these two properties is difficult but significant progress has been made in the last years, using sophisticated modelling of neutron star atmospheres and statistical analyses \cite{Steiner13}. However, definite conclusions are still controversial.

In this situation heavy ion collisions (HIC) from Fermi energies up to intermediate energies of several GeV per particle provide a means to investigate the nuclear EoS in the laboratory. In a collision nuclear matter first compresses and then expands in the final state, and thus different regions of density are explored. The advantage of heavy ion collisions is the freedom to vary the energy and the impact parameter (and thus the compression), and also the asymmetry of the colliding system within limits, which will be extended in the future with new rare isotope facilities. The difficulty is that a HIC is fundamentally a non-equilibrium process, and thus the detailed evolution has to be modelled using transport theory, on which there are still open questions. Also one has to identify observables which are especially sensitive to the symmetry energy in the presence of uncertainities of the much larger contribution of symmetric nuclear matter and the above-mentioned uncertainities of transport theory. 

Here I aim to give a brief overview of the present status of the investigation of the symmetry energy using HIC, using not only results of my collaborations but also those of other groups. Certainly this is not possible in any complete way in the limited space of this article. Thus I will select specific examples, which also point to the still open problems in these investigations. Recently there have been extensive reviews on the symmetry energy in the form of volumes collecting articles of experts \cite{SE_book06,SE_book14}, review articles \cite{Baran05,BALi08}, and feature articles \cite{Tsang12,Horow14}, just to mention the more recent ones, which contain much more information.

 \subsection*{2 Theoretical Considerations}  


The main method to interpret HIC is a transport theory, which describes the temporal evolution of the one-body
phase-space distribution function $f({\bf r,p},t)$ under the action of a mean
field potential $U({\bf r,p}),$ possibly momentum dependent, and 2-body
collisions with the in-medium cross
section $\sigma(\Omega)$. In a non-relativistic approach it reads

\begin{eqnarray}
\frac{df_i}{dt}  & = & \frac{\partial f_i}{\partial t}+\frac{\bf p_i}{m}\nabla^{(r_)}f_i
-  \nabla^{(r)}U_i({\bf r,p})\nabla^{(p)} f_i 
-\nabla^{(p)}U_i({\bf r,p})\nabla^{(r)} f_i \nonumber \\
& = & \sum_{j,i',j'}\int d{\bf p_j}d{\bf p_{i'}}d{\bf p_{j'}}
v_{ij}\sigma_{i,j\rightarrow i',j'}(\Omega)\delta({\bf p_i}+{\bf p_j}-{\bf p_{i'}}-{\bf p_{j'}}) \nonumber \\
&  & \times[(1-f_i)(1-f_j)f_{i'}f_{j'}-f_if_j(1-f_{i'})(1-f_{j'})], \nonumber 
\end{eqnarray}
but field theoretical formulations are also widely used \cite{Baran05}. The indices $(i,j,i',j')$ run over neutrons and protons, such that these are coupled equations via the collision term and indirectly via the potentials. If the production of other particles is considered, like $\Delta$'s, or $\pi$ and $K$ mesons, these have their own transport equations coupled through the corresponding inelastic cross sections. The mean field potentials $U_i$ can be derived from an energy functional. Mean fields and cross sections should be related through a theory for the in-medium effective interaction, like e.g. Brueckner theory, even though this is not necessarily done in many applications. The isospin effects enter via the differences in neutron and proton potentials and the isospin dependent cross sections, but they are always small relative to the dominant isoscalar effects. Thus one often resorts to differences or ratios of observables between isospin partners, in order to eliminate as much as possible the uncertainties in the isoscalar part. The decription of cluster production, which as seen below is often an issue in the investigations of the symmetry energy, in principle goes beyond the one-body description, and its proper treatment is still one of the questions in the application of transport theory.

\subsection*{3 Isospin observables}

It may be useful to briefly summarize the different ways in the investigations of the symmetry energy in the different density regimes.
In central reactions at Fermi energies densities somewhat above saturation are reached. Recently the expansion phase of such reactions has been studued in detail, where very low densities of about 1/10 to 1/1000 of $\rho_0$ are attained. From isostope ratios (so-called iso-scaling) the symmetry energy at very low densities has been determined
 which is important for the simulation for supernova explosions. In this density regime few-body clustering effects become important. A theoretical investigation has shown that the  symmetry energy, in fact, is finite a very low densities in qualitative agreement with experiment \cite{Natow10}.

At Fermi energy collisions one observes the phenomenon of multifragmentation. The distribution of the isospin to the different fragments ("isospin fractionation") \cite{Baran05}. and the isospin transport through the low-density neck ("isospin diffusion") in more peripheral collisions have  been very useful to constrain the symmetry energy below $\rho_0$ \cite{Tsang09}. 

At intermediate energies, which are the main interest here,
the initial phase of the collision is characterized by pre-equilibrium emission of high energy particles and light fragments. The yield ratios of isotopic
partners, like $n/p$ or $^3He/t$, contain information on the relative strengths
of the neutron and proton potentials. The compression phase then determines the momentum distribution of the emitted particles, generally called "flow", in-plane (directed) and out-of-plane (elliptic).
Neutron-proton differences of flow observables have been
an important means to extract information on the high density symmetry energy. Inelastic NN collisions lead to the production of $\Delta$ resonances, which may decay into pions or lead to the production of strangeness.

Let us briefly note that the influence of the symmetry energy has also been discussed at higher energies, suggesting that the deconfinement transition may be substantially influenced by the difference of the symmetry energy between the hadronic and partonic phases, and may, in fact, occur at lower density in asymmetric systems \cite{Shao11}.

\subsubsection*{3.1 Pre-equilibrium Emission}

\begin{figure}[t]
\begin{center}
\includegraphics[width=0.6\textwidth]{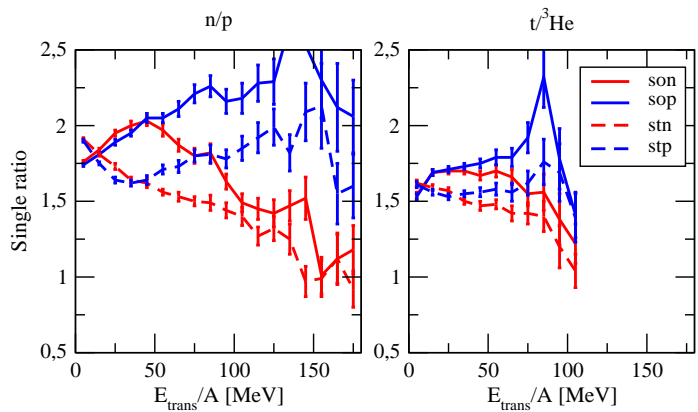}
\end{center}
\caption{(left) The neutron-proton ratio in $^{136}Xe+^{124}Sn$ collisions at 150 AMeV for different choices of the symmery energy (solid asy-soft, dashed asy-stiff) and orderings of the effective masses (blue, $m^*_n<m^*_p$, red $m^*_n>m^*_p$). On the right the corresponding ratio of tritium over $^3He$.}
\label{fig:fig_np}
\end{figure}

The neutron to proton ratio of emitted particles  has first been measured at MSU for $Sn+Sn$ systems at 50 AMeV, and a systematic analysis of several observables has yielded rather good limits on the $\gamma$ exponent around $\gamma\approx 0.6$\cite{Tsang09}. In the pre-equilibrium emission at higher energies the momentum dependence of the symmetry potential, i.e. the proton-neutron effective mass splitting, becomes important as first pointed out in refs. \cite{Rizzo05,Giord10}. Recently we have systematically studied this effect for nucleons and light clusters in different $Xe+Sn$ reactions at energies between 32 and 150 AMeV in ref.\cite{Wolter_INPC14}. A result from these calculations is shown in Fig.\ref{fig:fig_np} for central collisions at 150 AMeV with different stiffnesses of the symmetry energy and different effective mass splittings. in the left panel the $n/p$ ratio is shown as a function of the transverse energy of the emitted particles, on the right the corresponding result for $t/^3He$. One observes a clear pattern, namely that the stiffness of the symmetry energy governs the lower part of the transverse energy spectrum, where a softer symmetry energy yields a larger $n/p$ ratio. The higher energy part of the spectrum is dominated by the effective mass ordering, where a smaller neutron effective mass favors the emission of neutrons and increases the ratio. A similar result has been obtained by Zhang et al.\cite{Zhang14}. Thus this observable should serve as a promising probe to disentangle the density and momentum dependences of the symmetry potential at higher density. The ratio $t/^3He$ in the right panel shows a very similar pattern. A not yet very  conclusice comparison with INDRA data \cite{Wolter_INPC14} favors a stiff symmetry energy with  $m^*_n>m^*_p$.

\subsubsection*{3.2 Flow}

\begin{figure}[t]
\centering
\includegraphics[width=8cm]{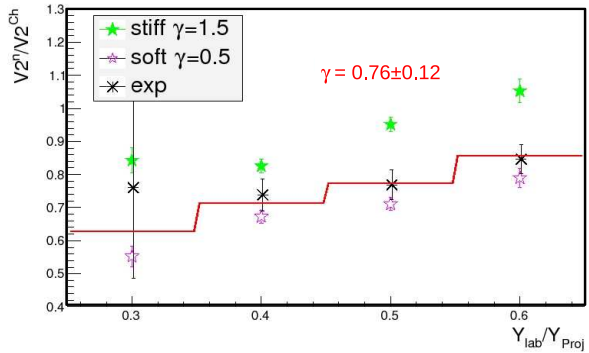}
\caption{ Ratio of elliptic flow of neutrons over hadrogen for forward rapidities in Au+Au collisions at 400 AMeV. Data of ref.\cite{Russo13} and calculations with a symmetry energy characterized by exponent $\gamma$.}
\label{fig:flow}
\end{figure}

The momentum distribution of the particles and fragments emitted in the final stage of a HIC are characterized via a Fourier series expansion of the azimuthal distribution as 
$N(\Theta;p_t,y)=N_0(1-v_1(p_t,y)cos\Theta+v_2(p_t,y)cos2\Theta+\ldots$ The first two Fourier coefficients, depending on the transverse momentum $p_t$ and the longitudinal rapidity $y$, are called directed and elliptic flow, respectively. Differences of flow parameters between isospin partners directly reflect the isospin-dependent potentials and thus the symmetry energy. Preliminary results from the ASYEOS experiment of the FOPI collaboration are shown in Fig.\ref{fig:flow} for Au+Au collisions at 400 AMeV \cite{Russo13}. Data for the ratio of the elliptic flow of neutrons relative to hydrogen are shown together with calculations for a soft and a stiff symmetry energy. A best fit yields a coefficient $\gamma$ of about 0.76, i.e. a moderately soft symmetry energy. This is an important advance in trying to constrain the symmetry energy at supersaturation densities.
 
\subsubsection*{3.3 Particle Production}

The $n/p$ asymmetry of the compressed system also influences the ratio of newly produced particles, which thus can serve as indicators of the symmetry energy in the high density  phase. Pions are produced predominantly via the $\Delta$ resonances, $NN\rightarrow N\Delta$ and the subsequent decay $\Delta\rightarrow N\pi$. The ratio of the isospin partners $\pi^-/\pi^+$ can thus serve as a probe of the high density symmetry energy. As analyzed in ref.\cite{Ferini05} there are competing effects on the $\Delta$ and pion production from the isospin dependent mean fields and the $\Delta$ production threshold conditions.

\begin{figure}[t]
\begin{center}
\includegraphics[width=13cm]{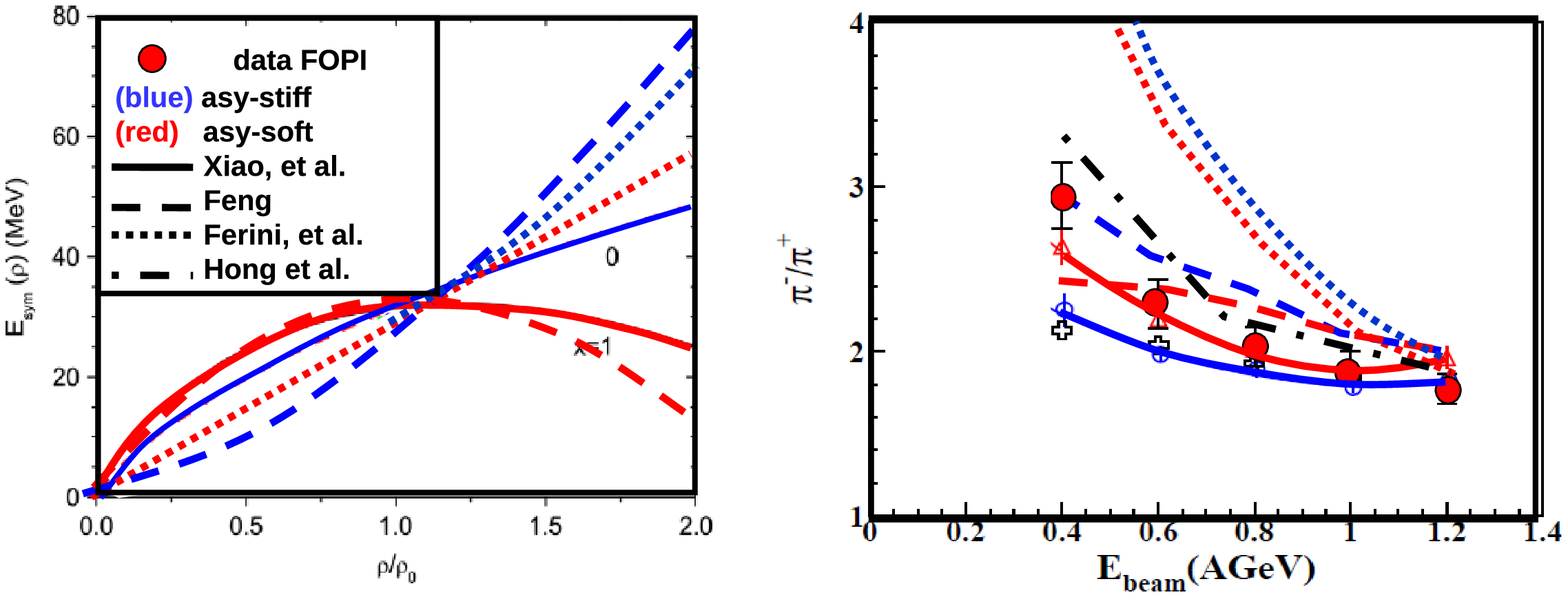}
\end{center}
\caption{ (right) The $\pi^-/\pi^+$ ratio in $Au+Au$ collisons as a function of incident energy as measured by the FOPI collaboration, and as calculated by different groups, as indicated in the legend and discussed in the text. On the left are shown the corresponding models for the symmetry energy.}
\label{fig:pions}
\end{figure}

In Fig.\ref{fig:pions} I have collected in the right panel results from recent theoretical analyses of this ratio using different models of symmetry energies, shown correspondingly in the right panel \cite{pions}, and different program codes. They are compared to the FOPI data \cite{Reisd07}. For each model the results for two parameter sets of different stiffness are shown (stiffer - blue, softer - red). As is seen, the results of the different models are not only very different quantitatively, but even the trend with the asy-stiffness is not consistent.
A reason may lie in different modelling of the $\Delta$ dynamics, and also in the competing  mean field and the threshold effects, where slightly different treatments might lead to large differences.
This issue needs clarification in view of the sensitivity  of the pion observables and the excellent data situation. More information should be gained by discussing not the energy integrated yield ratios, but the spectral behavior of the ratio, since different energy pions are expected to have different histories \cite{Lynch15}.

It has also been suggested, that the ratio of the anti-strange kaon isospin partners, $K^0/K^+$ could be a useful observable for the symmetry energy \cite{Ferini06}. Indeed, kaon production has been one of the most useful observables to determine the EOS of symmetric nuclear matter. The anti-strange kaons  weakly interact with nuclear matter and are thus a direct probe of the dense matter where they are produced. Theoretical analyses show similar if not larger sensitivity to the symmetry energy compared to pion ratios, but appropriate date do not yet exist. 

\begin{figure}
\begin{center}
 \parbox{6.5cm}{\includegraphics[width=5.5cm]{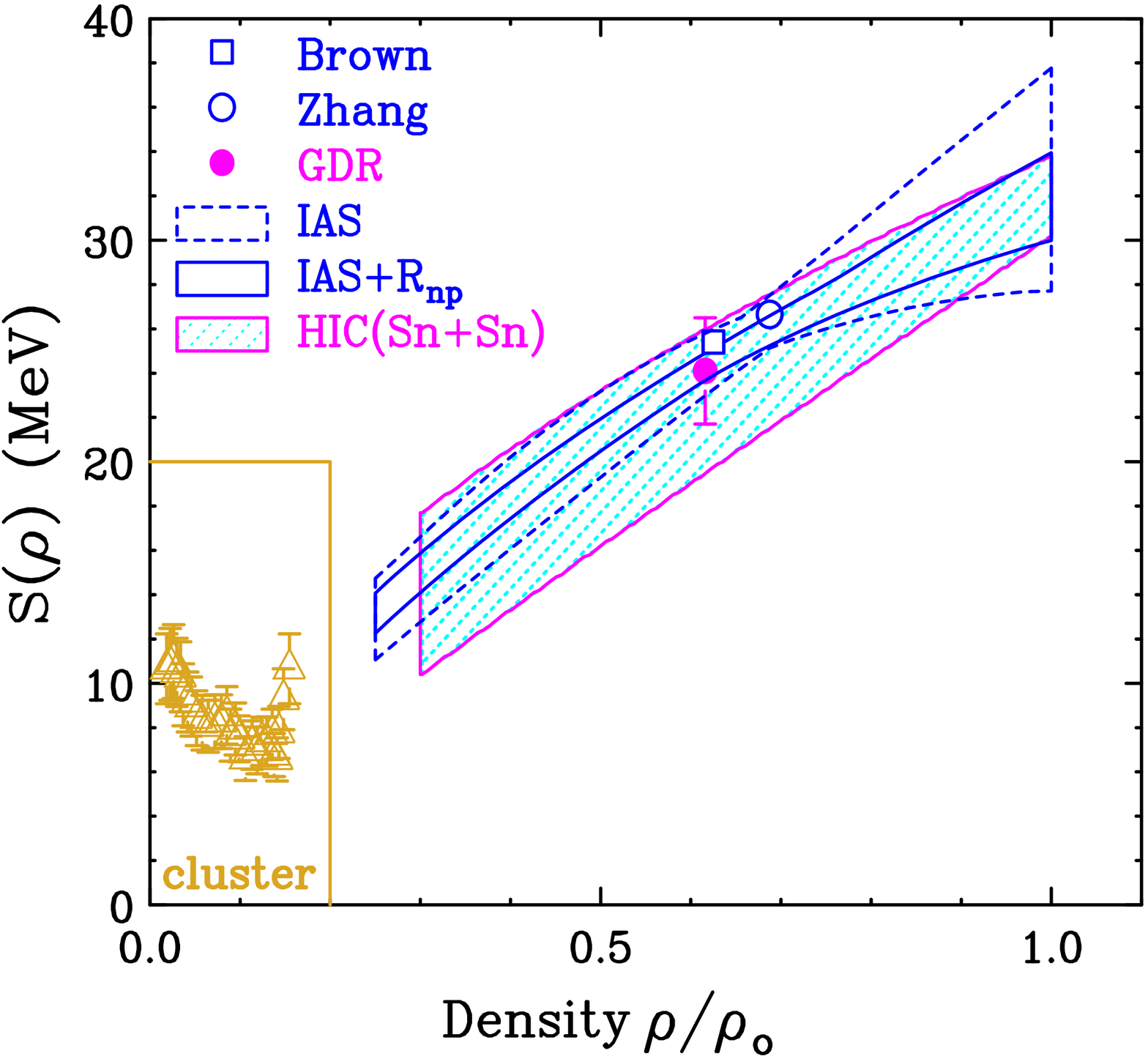}}
 \hspace*{0.3cm}
 \parbox{7.0cm}{\includegraphics[width=7.0cm]{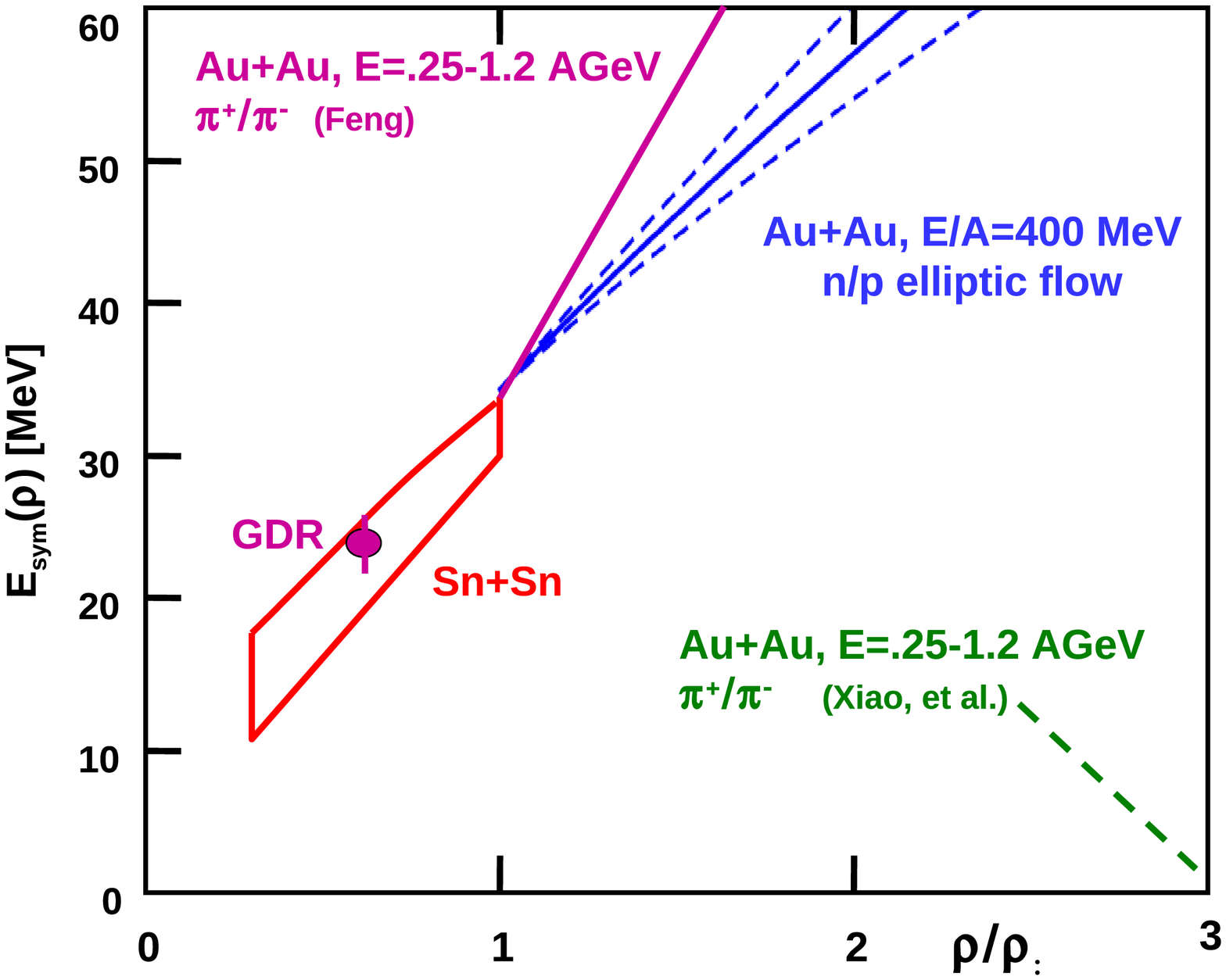}}
 \caption{Summary of information on the density dependence of symmetry energy for below (left) and above (right) saturation density. On the left information is collected from nuclear structure and low energy HIC \cite{Horow14}, while on the right the still not consistent information from higher energy collisions. The figure is discussed in the text. }
\label{fig:Esym}
\end{center}
\end{figure}

\subsection*{4 Discussion and Conclusions}

In Fig.\ref{fig:Esym} I attempt to give a summary of the present information on the density dependence symmetry energy $E_{sym}(\rho)$ or equivalently $S(\rho)$ from HIC \cite{Tsang12,Horow14}. In the left panel the region below saturation density $\rho_0$ is shown. The blue hatched area is the result from the investigation of Sn+Sn collisions at 50 MeV from MSU \cite{Tsang12} using various observables from isospin transport between different Sn isotopes. The isolated symbols represents information from the fits to nuclear masses or GDR energies, which are plotted at about $0.6\rho_0$, which is an average density of nuclei where different models of the EoS converge. The blue-bordered areas are derived from an analysis of isobaric analog states which give information also on the lower densities in the surface \cite{Danielew13}. When the analysis is combined with the information on the neutron skin radius of Pb, the constraint is still sharpened. The points in the lower left corner, labelled "cluster", come from an analysis of the very low density matter in the expansion phase of low energy heavy ion collisions, mentioned above \cite{Typel10}. Here the matter is not any more homogeneous but cluster correlations become important, which have the effect of making the symmetry energy finite at very low densitiy. 
Alltogether the various sources of information on the symmetry energy in this density range seem to converge, and they also converge with the theoretical many-body results, see e.g. ref.\cite{Fuchs06}.

The information on the symmetry energy above saturation is shown in the right panel (where the low densitiy results from the right panel are shown  by the red-bordered area),  particularly the results from the neutron/hydrogen flow analysis from sect. 3.3, and two results from the analysis of the pion ratios from sect. 3.4, one favoring a very soft symmetry energy and the other a rather stiff one \cite{pions}. Microscopic many-body results in the region up to $2\rho_0$ favor a behavior more similar to the flow experiment. 

I have attempted to give a brief overview of the determination of the nuclear symmetry energy in HIC's.
HIC's are interpreted with transport theories and I have mentioned some of the
challenges in such descriptions. Today a
picture emerges where the information on the symmetry energy from HIC's, nuclear structure,
and neutron stars increasingly converges. But there are obviously also open problems. where a
more thorough understanding of the mechanism and the analysis is needed. In the end it is
neccessary to obtain a consistent picture for many observables in heavy ion
collisions.

\subsection*{Acknowledgement}

This article has benefitted greatly from the work both of a long-standing collaboration with the Catania reaction group consisting of, among others, Maria Colonna, Massimo Di Toro, Vincenzo Greco, Theo Gaitanos (now Thessaloniki), Malgorzata Zielinska-Pfabe and Pjotr Decowski (recently deceased) (both Smith College, USA), and from the compact star group, consisting of, among others, Gerd R\"opke (Rostock), David Blaschke and Thomas Kl\"ahn (Wroclaw) and Stefan Typel (GSI). I also want to thank many colleagues for valuable discussions, in particular W.Trautmann (GSI), Betty Tsang and Pawel Danielewicz (MSU). The work is supported by the German Science Foundation cluster of excellence {\it Origin and Structure of the Unverse}.

\end{document}